# Ultrasound-Mediated Cavitation-Enhanced Extravasation of Mesoporous Silica Nanoparticles for Controlled-Release Drug Delivery

Juan L. Paris[1,2,+], Christophoros Mannaris[3,+], M. Victoria Cabañas[1], Robert Carlisle[3], Miguel Manzano[1,2], María Vallet-Regí[1,2], Constantin C. Coussios[3,*].

1. Dpto. Química Inorgánica y Bioinorgánica, Facultad de Farmacia, UCM, Instituto de Investigación Sanitaria Hospital, 12 de Octubre i+12, 28040-Madrid, Spain.

2. Centro de Investigación Biomédica en Red de Bioingeniería, Biomateriales y Nanomedicina (CIBER-BBN), Spain.

3. Institute of Biomedical Engineering, University of Oxford, Old Road Campus Research Building, Oxford OX3 7DQ, UK.

[+]Authors contributed equally to the work

*Corresponding author: constantin.coussios@eng.ox.ac.uk

## Abstract

Mesoporous silica nanoparticles have been reported as suitable drug carriers, but their successful delivery to target tissues following systemic administration remains a challenge. In the present work, ultrasound-induced inertial cavitation was evaluated as a mechanism to promote their extravasation in a flow-through tissue-mimicking agarose phantom. Two different ultrasound frequencies, 0.5 or 1.6 MHz, with pressures in the range 0.5-4 MPa were used to drive cavitation activity which was detected in real time. The optimal ultrasound conditions identified were employed to deliver dye-loaded nanoparticles as a model for drug-loaded nanocarriers, with the level of extravasation evaluated by fluorescence microscopy. The same nanoparticles were then co-injected with submicrometric polymeric cavitation nuclei as a means to promote cavitation activity and decrease the required in-situ acoustic pressure required to attain extravasation. The overall cavitation energy and penetration of the combination was compared to mesoporous silica nanoparticles alone. The results of the present work suggest that combining mesoporous silica nanocarriers and submcrometric cavitation nuclei may help enhance the extravasation of the nanocarrier, thus enabling subsequent sustained drug release to happen from those particles already embedded in the tumour tissue.

**Introduction**

Cancer remains one of the leading causes of death in the world.[1] The poor delivery and inadequate distribution of cancer therapeutics in tumours is considered to be one of the major hurdles yet to be overcome.[2] The use of nanocarriers to transport anticancer drugs has been extensively studied in recent years as a promising strategy to address this challenge.[3] Drug nanocarriers can provide many advantages, such as improving the pharmacokinetic profile of drugs, increasing their circulation times and improving their safety profile.[4] Mesoporous Silica Nanoparticles (MSNs) are one of the most promising types of nanoparticles for drug delivery, due to their high drug-loading capacity derived from their textural properties, such as their high surface area and pore volume, their physicochemical robustness and their ease of functionalization with different moieties through silanol chemistry, making them a versatile tool that can be easily adapted to different specific applications.[5,6]

The main rationale behind the vast amount of work in nanomedicine is the selective accumulation of nanostructures in solid tumours due to their anomalous vasculature related to their rapid growth, which constitutes the enhanced permeation and retention effect, often referred to as "passive targeting".[7] Despite all the efforts spent in developing the field of nanomedicine, only a few nano-sized drugs have reached clinical application.[8] Furthermore, while these clinically available nano-drugs have significantly improved the safety profile compared to previous formulations with the same active components, mainly due to preventing the use of toxic excipients, the improvement in their efficacy has been modest at best.[9] A recent analysis on the accumulation of nanoparticles in tumours has shown that only a small percentage of the injected nanocarriers actually reach the target tumour tissue (*ca.* 0.7 %).[10] Besides, there has been no significant improvement in that parameter over the last decades.[10] Moreover, the heterogeneity of tumours in different patients, or even on different areas of the tumours from the same patient, implies that even the nanoparticles that reach them will not be distributed homogeneously. Furthermore, the elevated interstitial pressure of solid tumours also limits the penetration of nanoparticles into the tissue.[11] Consequently, a more controlled method to deliver cancer nanotherapeutics is necessary, while also ensuring a deeper penetration of the nanoformulations to allow them to achieve their full therapeutic potential.

Ultrasound (US) has emerged as a remarkable tool in biomedicine, being used to induce drug release from drug carriers,[12–16] to perform thermal ablation therapies[17] and to induce reversible opening of the blood-brain barrier,[18] among many other applications.[19–21] Ultrasound can be focused in the region of interest deep within the body, reducing the intensity of the stimulus applied to surrounding, non-target areas. The biological effects of ultrasound can be divided into thermal and mechanical.[17] Among those mechanical effects, cavitation is highlighted as one of the most useful ultrasound-related phenomena for biomedicine. It can be defined as the formation and linear or non-linear oscillation of gas bubbles in a fluid.[22,23] Sustained linear or non-linear oscillations about an equilibrium bubble radius for many acoustic cycles is termed non-inertial cavitation. However, at more elevated pressures for a given ultrasound frequency and bubble size, the bubble grows unstably during ultrasonic rarefaction, subsequently collapsing violently during compression under the inertia of the surrounding fluid.

This phenomenon, termed inertial cavitation (IC), has been used to enhance the delivery of different therapeutics for cancer therapy.[22,24–26] In this sense, the main strategy employed here consists on using cavitation nuclei that reduce the acoustic pressure required to induce IC, which implies violent collapse of gas bubbles.[27] Examples of such nuclei include micrometric shelled gas bubbles,[22] droplets [28] or gas-stabilizing solid nanoparticles.[29,30] Microstreaming, shockwaves and microjets, associated with IC within blood vessels, can propel the therapeutic into the tumour, increasing the delivered dosage and favouring a more homogeneous distribution.[31]

The main objective of the present work is to evaluate in an *in vitro* tissue mimicking flow model the possibility of MSNs to be used both as drug carriers and as nuclei for the generation of IC to improve their delivery and penetration. Additionally, the combination of MSNs with sub-micron polymeric cup-shaped gas-stabilizing sonosensitive particles (SSP)[29,30] to reduce the acoustic pressures required to initiate IC is also investigated.

**Experimental section**

*Materials.* Following compounds were purchased from Sigma-Aldrich Inc.: Aminopropyltriethoxysilane (APTES), ammonium nitrate, cetyltrimethylammonium bromide (CTAB), sodium hydroxide, tetraethyl orthosilicate (TEOS), fluorescein isothiocyanate (FITC), phosphate-buffered solution (PBS), rhodamine B. These compounds were used without further purification. UltraPure™ Agarose-1000 (Invitrogen, Paisley, UK) was also used without further purification. Ultrapure deionized water was obtained using a Millipore Milli-Q plus system (Millipore S.A.S., France).

*Characterisation Techniques.* The materials were analysed by small angle X-ray diffraction (XRD) in a Philips X'Pert MPD diffractometer equipped with Cu Kα radiation. Fourier transform infrared (FTIR) spectra were obtained in a Nicolet (Thermo Fisher Scientific) Nexus spectrometer equipped with a Smart Golden Gate Attenuated Total Reflectance accessory. Transmission electron microscopy (TEM) was carried out in a JEOL JEM 2100 instrument operated at 200 kV, equipped with a CCD camera (KeenView Camera). The zeta potential and hydrodynamic size of nanoparticles determined by dynamic light scattering (DLS) were measured by means of a Zetasizer Nano ZS (Malvern Instruments) equipped with a 633 nm "red" laser. $N_2$ adsorption was carried out in a Micromeritics ASAP 2010 instrument; surface area was obtained by applying the Brunauer–Emmett–Teller (BET) method to the isotherm and the pore size distribution was determined by the Barrett–Joyner–Halenda (BJH) method from the desorption branch of the isotherm. Mesopore diameter was obtained from the maximum of the pore size distribution curve.

Fluorescence spectrometry was used to determine cargo release by means of a Biotek Synergy 4 device ($\lambda_{exc}$ 540 nm, $\lambda_{em}$ 625 nm). Fluorescence microscopy was performed in a Nikon Super Inverted Research Microscope (Ti-E). Individual images were taken using a 4x objective. 3D image stitching was performed using the software NIS-Elements Advanced Research (Nikon UK Limited). Nikon filter cubes were used to image FITC (filter cube for FITC, Exciter 465–495, Dichroic 505, Emitter 515–555) and rhodamine B (filter cube for TRITC, Exciter HQ545/30×, Dichroic Q570LP, Emitter HQ620/60 m).

*Synthesis of Mesoporous Silica Nanoparticles*: MSNs were prepared by a modified Stöber method, as previously described.[32] Briefly, 1 g of the surfactant CTAB was dissolved in 480 mL of deionised (DI) water with 3.5 mL of NaOH 2M in a 1 L round-bottom flask. The solution was heated up to 80 ºC under magnetic stirring, and the temperature was allowed to stabilize for 30 min. 5 mL of TEOS were added at 0.25 mL/min and the solution was vigorously stirred at 80 ºC for 2 h. The particles were then collected by centrifugation and washed with water and ethanol. The surfactant was then extracted by ionic exchange by dispersing the particles in 500 mL of $NH_4NO_3$ solution in 95% ethanol (10 mg/mL) and magnetically stirring them under reflux overnight. The nanoparticles were then collected by centrifugation and washed twice with ethanol. The resulting solid was dried under vacuum at room temperature.

FITC-labelled MSNs (FMSNs) were prepared in a similar manner, by conjugating 1 mg of FITC with 2.2 µL of APTES in 100 µL of ethanol with magnetic stirring for 2 h. That solution was mixed with 5 mL of TEOS and added onto the CTAB solution under basic conditions, following the same protocol as described above.

*Loading and release of Rhodamine B in FMSNs*: 2 mg of Rhodamine B were dissolved in 20 mL of deionised (DI) water. 20 mg of FMSNs were then dispersed in the dye solution with the help of an ultrasound bath and the suspension was magnetically stirred at room temperature overnight. The dye-loaded particles (FMSN-RhB) were then collected by centrifugation and thoroughly washed with DI water before being dried and stored at room temperature until further use.

Rhodamine B release experiments were carried out as follows. A 1 mg/mL suspension of FMSN-RhB in water was prepared and 0.5 mL of such suspension were introduced in a Transwell® insert (polycarbonate membrane with a pore size of 0.4 µm) placed in a 12 well plate. 1.5 mL of DI water were placed in the well outside the insert. At predetermined time points, the release media outside the inserts were collected and replaced with fresh DI water. The samples were analysed by fluorimetry using a plate reader. Release experiments were performed in triplicate.

*Sonosensitive Particles (SSPs):* Submicrometric polymeric cups were obtained from OxSonics Ltd (Oxford, UK).[30] Degassed DI water was used to dilute a stock solution of 25 mg/mL of SSPs to a final concentration of 0.5 mg/mL. The SSPs had a mean diameter of 500 nm, with a cavity diameter between 230 and 340 nm.[30] Suspensions of both SSPs and FMSNs combined (final concentration of 0.5 mg/mL and 0.2mg/mL, respectively) were also prepared employing degassed DI water (FMSN-SSP suspension).

*Experimental setup:* The experimental setup used in this work had four main components: a focused therapeutic ultrasound transducer (FUS), a passive cavitation detector (PCD) that is used to passively record the acoustic emissions produced from cavitation, the *in vitro* tissue-mimicking agarose phantom model (with an embedded 1 mm channel through which the sample can flow) and a conventional ultrasound imaging device. The FUS and PCD are fully controlled from custom-made software using graphical programming language (LabVIEW, National Instruments, USA). An arbitrary waveform generator (33220A, Agilent, USA) was used to create the transmit signal which was amplified by a 300W RF power amplifier (A-300, ENI, USA) and sent to the FUS via a 50 ohm matching network.

   *1. Focused therapeutic ultrasound:* A spherically-focused single-element FUS transducer with a centre frequency of 0.5 MHz was used (H107, Sonic Concepts, USA). The same transducer driven at its third harmonic through a customized matching network was used for the 1.6 MHz experiments. The aperture and the geometric focus of the transducer were 64 mm and 60 mm, respectively. The FUS used was previously calibrated in water using a 0.4 mm diameter needle hydrophone (ONDA 1056, Onda Corporation, USA). All acoustic pressures reported in this study are in MPa peak rarefactional pressures (PRP). For all extravasation experiments, 600 pulses with a pulse repetition time (PRT) of 100 ms were employed. Duty cycle was kept at 5 % in all cases.

*2. Passive Cavitation Detector:* A single element PCD (V320, Panametrics, USA) was coaxially and confocally aligned with the FUS transducer via a central circular opening in the FUS. Acoustic emissions arising from cavitation were sensed by the PCD as described previously.[33] Here, a 7.5 MHz spherically focused PCD of element diameter 12.5 mm and focal length 75 mm was used. The acquired PCD signal was filtered using a 5 MHz high pass filter (FILT-HP5-A, Allen Avionics), amplified 5 times with a low noise amplifier (Stanford Research Systems, SR445A) and recorded with a 14-bit PCI Oscilloscope device (PCI-5122, National Instruments, USA) at a rate of 100 MHz. The high-pass filter was used to reject strong reflections from the agarose phantom at the fundamental frequency (and harmonics due to non-linear propagation of the incident beam). The pressure threshold required to initiate cavitation activity was determined by 50-cycle FUS excitation pulses that were ramped at small pressure increments under constant flow. A typical PCD data trace consisted of an initial ~85μs segment ('background') that was free of any signal content in the filter pass band, followed by scattering and cavitation emissions ('signal') whose durations varied with drive pulse length. The background and signal segments of each trace were analysed in MATLAB (Mathworks, USA) to determine if IC had occurred and the full ensemble of PCD traces were reviewed to calculate a probability of cavitation at the prescribed ultrasound settings. For this analysis, any harmonic components were removed by post-processing with a digital comb filter. The identification of IC was based upon broadband spectral elevation in the signal relative to the background. The harmonic-suppressed traces were deemed to exhibit IC when the mean-squared signal was at least 10 times larger than the background signal.

*3. Tissue-mimicking agarose phantom:* A degassed biocompatible hydrogel composed of 1.25 % (w/v) low melting point ultrapure agarose gel with an embedded 1 mm channel was created by heating and cooling process.[22] The phantom contained three 50 mm long channels which allowed multiple conditions to be tested in the same gel, thus minimising variability. A clear and acoustically transparent Mylar film was used to isolate the gel from the surrounding water and allow free propagation of ultrasound. A low-pulsatility peristaltic pump (Minipulse Evolution, Gilson, USA) was used to flow the FMSNs or FMSN-SSP suspension at a constant flow rate of 0.2 mL/min. The flow rate was chosen to avoid channel rupture and leakage while still in agreement with previously published data of tumour perfusion.[34]

*4. Ultrasound imaging:* A Philips iU-22 ultrasound scanner was used to provide real-time imaging of the flow phantom during the experiment. A linear L12-5 probe was placed at an angle of 60° to the FUS and PCD propagation axis and imaging was done at a low mechanical index (MI < 0.05), in order to minimise interference with the experiment. The imaging allowed monitoring of the channel and its contents during the experiments and provided real-time feedback of the therapeutic process.

*Experimental Procedure:* All experiments were carried out in a large tank filled with DI water that was degassed overnight prior to each experiment. With the channel filled with air, the FUS was driven in pulse-echo mode using a pulser-receiver (DPR300, JSR Ultrasonics, USA) in order to ensure that the FUS and PCD were aligned with the middle of the flow channel. The sample was then introduced at a constant flow rate and was kept flowing during FUS exposures. Figure 1A shows a schematic representation of the experimental setup. The input voltages to the FUS matching network were also recorded and converted to peak negative pressures using previous hydrophone calibrations. Focused ultrasound was applied at different positions within the channel, while monitoring both the passive cavitation detector data (to detect cavitation during the ultrasound exposure) and performing B-mode imaging simultaneously. A photograph of the agarose phantom holder and examples of B-mode images taken during the experiments can be seen in Figure 1B and 1C, respectively.

Flow channels were flushed with DI water following ultrasound exposure to remove any remaining fluorescent particles. A rectangular prism containing the flow channel, approximately 40 mm long, was then excised and placed on a glass microscope slide to evaluate nanoparticle extravasation through microscopy (extravasation is here defined as any particle fluorescence detected outside the flow channel).

**Results and Discussion**

Nanoparticle delivery experiments were carried out in an *in vitro* agarose phantom model. The agarose gel acts as tumour tissue mimicking material, having a porosity in the same range (*ca.* 500 nm pore size) as the endothelial gaps of the "leaky" tumour tissue.[35]

Cavitation has been previously used to induce the extravasation of nanoparticles to tumour tissues.[22,24–26] The main objective of this work is to evaluate *in vitro* the extravasation of MSNs in the presence of acoustic cavitation with or without SSPs. Previously dried FITC-labelled MSNs (FMSNs) were dispersed in degassed water at a concentration of 200 µg/mL. FMSN suspensions were prepared by placing them in an ultrasound bath for 10 seconds 3 times, with gentle stirring after each time. The cavitation threshold of the FMSNs suspension under flow through the agarose phantom was evaluated at two different frequencies (0.5 and 1.6 MHz) (Figure 2). The duty cycle (DC) in both cases was limited to 5 % in order to minimise ultrasound-induced hyperthermia. IC could be detected at both frequencies with those FMSN suspensions (and no cavitation could be detected under the applied pressures at any frequency with just degassed water). At 0.5 MHz, cavitation was observed at pressures as low as 0.6 MPa, and a threshold was observed at 2 MPa where sustained cavitation (probability > 0.7) is achieved. At 1.6 MHz, cavitation was first observed at 1.5 MPa and linearly increased to a probability of 0.71 at 4.3 MPa.

The IC detected with FMSN suspensions could be attributed to air trapped during drying process which could be on the surface or inside nanoparticle aggregates.[36] FMSN suspensions were evaluated by Dynamic Light Scattering (DLS) to check for nanoparticle aggregates (Figure S1). While no significant percentage of nanoparticle aggregates could be observed in the number percentage, some aggregates of several microns were detected in the intensity percentage (where larger particulates give more intense signals).

Having confirmed the onset of IC when exposing FMSN suspensions to focused US, the extravasation and delivery of those FMSNs in the agarose flow phantom model was evaluated next. FMSN suspensions flowing through the agarose phantom were exposed to focused ultrasound at 0.5 and 1.6 MHz, at two different PRP for each frequency (1 and 2 MPa or 2 and 4 MPa, respectively). The agarose phantoms were then cut and observed under the fluorescence microscope. Figure 3 shows a schematic representation of the nanoparticle delivery experiments and representative images of the results. Extravasated nanoparticles can be seen as green fluorescent spots in the agarose gel, outside the flow channel. As expected, the extravasation of FMSNs was more successful at higher pressures for each frequency, particularly once the cavitation threshold identified in Figure 2 was crossed. Differences were observed in the extravasation profiles, with 1.6 MHz achieving higher penetration depths but a more directional extravasation profile compared to 0.5 MHz, presumably due to the greater effect of acoustic radiation force at the higher frequency. The maximum penetration depth of FMSNs after focused US exposure under those conditions can be found in Figure S2.

The main rationale behind the use of MSNs in nanomedicine is the possibility of including drugs and other small molecules inside their pores to be then released in a controlled manner. Our next experiment consisted on evaluating nanoparticle delivery in the same model, but using loaded nanoparticles, to see if they could act as drug carriers while still being capable of propelling themselves into the tumour when insonified. FMSNs were therefore loaded with rhodamine B (FMSN-RhB). The successful loading can be observed by the colour change in the material, as well as by the FTIR spectrum and small angle XRD patterns (Figure S3). The drastic changes in BET surface area (from 1050 $m^2/g$ to 560 $m^2/g$) and pore volume further confirm that the mesopores of the material have been occupied by rhodamine B. TEM micrographs also show that the ordered porosity is maintained after cargo loading and nanoparticle washing. Release experiments of the dye from FMSN-RhB show a sustained release of the cargo into the aqueous medium (Figure S4). Rhodamine B release followed first order kinetics, with a faster initial release followed by a slower release that can last for several days, as previously observed for mesoporous drug release matrices.[12,37]

The cavitation activity of a suspension of FMSN-RhB exposed to 1.6 MHz focused ultrasound was also evaluated following the same procedure as had been performed for FMSN and is shown in Figure S5. Results are very similar to FMSN indicating that the loading of the dye did not affect the cavitation activity of the particle suspensions. The size distribution of FMSN-RhB evaluated by DLS was similar to FMSN, with some aggregates being detected only in the intensity percentage distribution (Figure S6). 1.6 MHz, 4 MPa was chosen for the FMSN-RhB delivery experiments, as the extravasation profiles observed at these settings were easier to detect under the microscope. Figure 4 shows the successful extravasation of the nanoparticles into the agarose gel (green fluorescence). The fluorescence of the cargo (red channel) can also be observed, both inside the nanoparticles (fluorescent spots colocalizing with the nanoparticles, in the green channel), and partially released from the material. This could be explained because the gels were cut and imaged shortly after delivery, so the model drug did not have enough time to be released and diffuse in the surrounding tissue, as would be the case in an *in vivo* scenario. Dye fluorescence can also be observed around the channel, due to the diffusion of fluorophore released during nanoparticle flow through the channel.

These results show the feasibility of using MSNs as carriers of small molecules that could be propelled by the application of focused ultrasound through the onset of IC. However, the pressures required to induce MSN extravasation and delivery with FUS might be too high for the clinical application of this strategy. The use of cavitation nuclei has gathered a lot of attention in recent years, since their usage can greatly decrease the pressure threshold needed to generate cavitation.[27,38] Among the different cavitation nuclei that have been used to enhance drug delivery, submicrometer sized particles hold great promise due to their capacity of extravasating along with the therapeutic, and further propelling the drug into the tissue of interest. Microbubbles on the other hand would be restrained in the vascular compartment and unable to extravasate along with the therapeutic being employed.[30] However, combining cavitation nuclei directly with an anticancer drug would only provide an initial dose of the drug, without establishing a stable concentration of the drug over time. For these reasons, we studied the combination of mesoporous silica nanoparticles with submicrometric polymeric cups. In this manner, a lower ultrasound pressure could be applied to extravasate MSNs, which would then act as a reservoir of the drug that would be released in the tumour area. Figure 5 shows the successful delivery of FMSNs into the agarose gel using 1.6 MHz FUS at just 2 MPa of pressure, a pressure which is readily achievable clinically without significant bioeffects and at which no extravasation had been previously observed in the absence of cavitation nuclei (Figure 3). The cavitation energy measured with the combination of FMSNs and SSPs greatly exceeded the energy reached at double the pressure without SSPs (Figure S7). Temperature elevation measurements were also performed with a fine wire thermocouple during FUS exposure (1.6 MHz, 4 MPa, 300 pulses, PRT of 100 ms) using both water or a SSPs suspension. The temperature increase produced under those conditions was negligible, below 1 ºC, indicating that thermal effects were unlikely to be the driving cause of any of the observed results (Figure S8). All of these results highlight the great potential of the combination of both types of particles for therapeutic application in drug delivery.

Finally, the combination of SSPs and FMSN-RhB shows significant delivery of FMSN-RhB to the agar gel, using 1.6 MHz US driven at 2 MPa PRP (Figure 6). Rhodamine B can be observed both still inside MSN and partially released into the surrounding gel. These results can be interpreted as a proof of concept of our approach, co-injecting SSPs with loaded MSN. A combination of both materials could be injected in a patient and US could be applied in the tumour area, activating SSPs and inducing IC in the area, propelling the particles towards the tumour tissue. Thus, drug-loaded MSN would be embedded in the tissue, which would then slowly release an anticancer drug close to its target cancer cells.

**Conclusions**

Ultrasound-induced inertial cavitation (IC) has been shown here to enable enhanced extravasation of mesoporous silica nanoparticles in an *in vitro* flow-through agarose tissue phantom. Both empty and dye-loaded nanoparticles were shown to extravasate into the agarose gel at ultrasound pressures beyond the IC threshold. The combination of those nanoparticles with submicrometric cavitation nuclei was demonstrated as a much more efficient way of inducing nanoparticle extravasation, decreasing the pressure needed to observe the desired extravasation effect by half. A two-stage strategy for the combination of both agents therefore emerges: in the initial step, the cavitation nuclei will drive nanoparticle extravasation and distribution throughout the tumour tissue when activated by focused ultrasound; mesoporous silica nanoparticles embedded in the target tissue then slowly release the drug of interest directly in the tumour mass.

**Acknowledgements**

The authors gratefully acknowledge funding from the European Research Council through the Advanced Grant VERDI (ERC-2015 AdG proposal no. 694160), Ministerio de Economía y Competitividad, (MEC), Spain (Project MAT2015- 64831-R), and the Oxford Centre for Drug Delivery Devices under programme grant EP/L024012/1 by the UK's Engineering and Physical Sciences Research Council.. JL Paris gratefully acknowledges MEC, Spain, for his PhD grants (BES-2013- 064182, EEBB-I-17-12352).

degree of SBA-15 as key factor to modulate sodium alendronate dosage, Microporous Mesoporous Mater. 116 (2008) 4–13.

[38] S. Datta, C.-C. Coussios, L.E. McAdory, J. Tan, T. Porter, G. De Courten-Myers, et al., Correlation of cavitation with ultrasound enhancement of thrombolysis, Ultrasound Med. Biol. 32 (2006) 1257–1267.

**Figure 1**

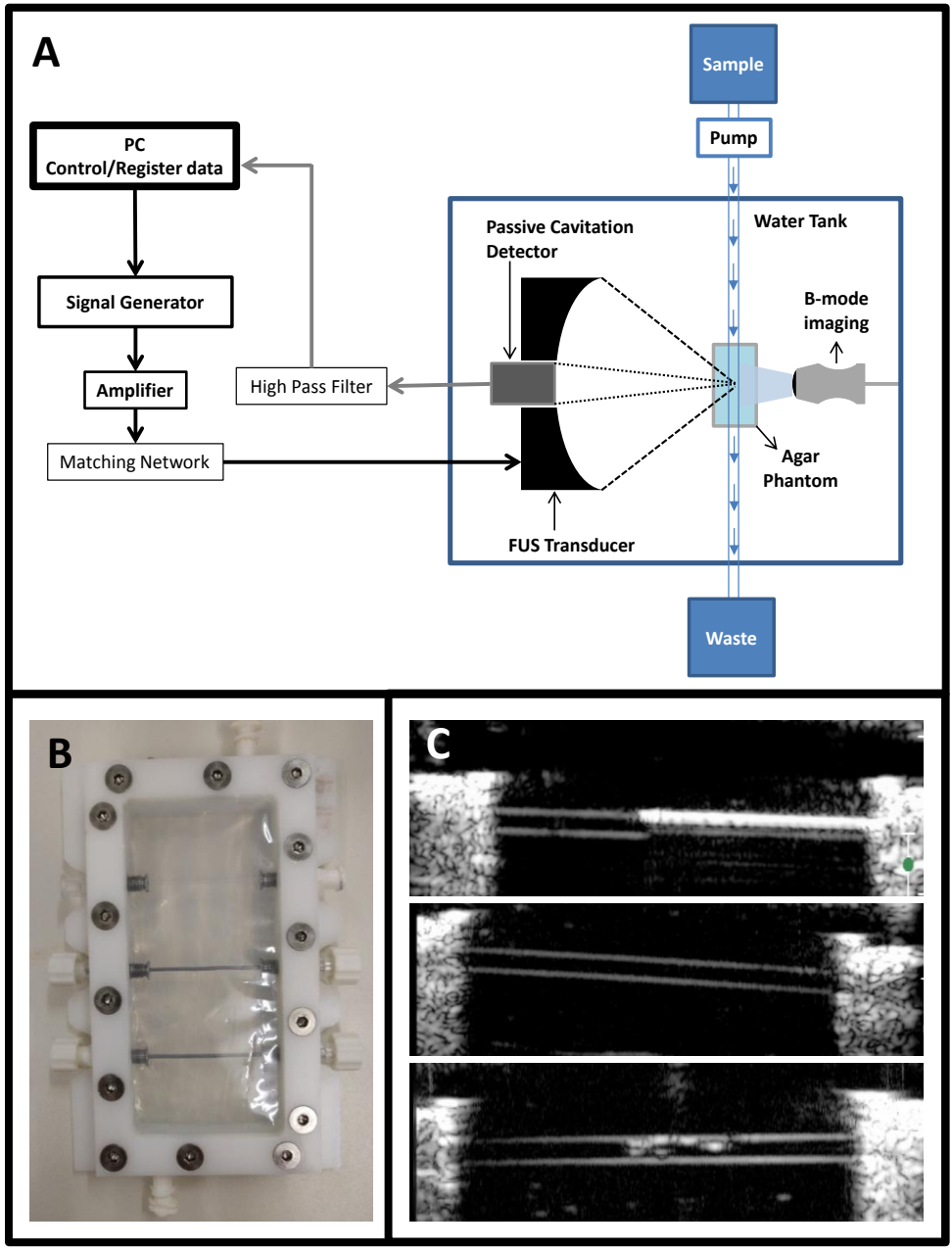

Figure 1.

**Figure 2**

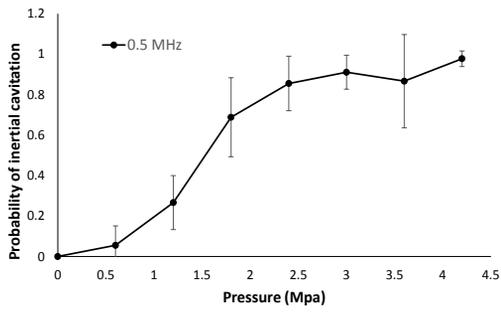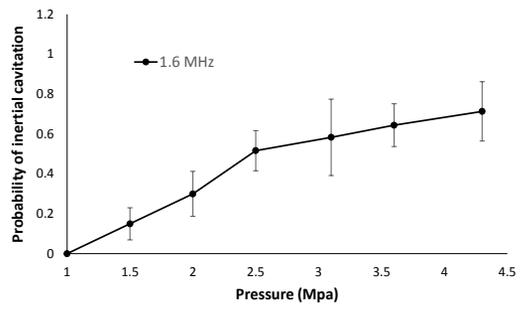

Figure 2.

Figure 3

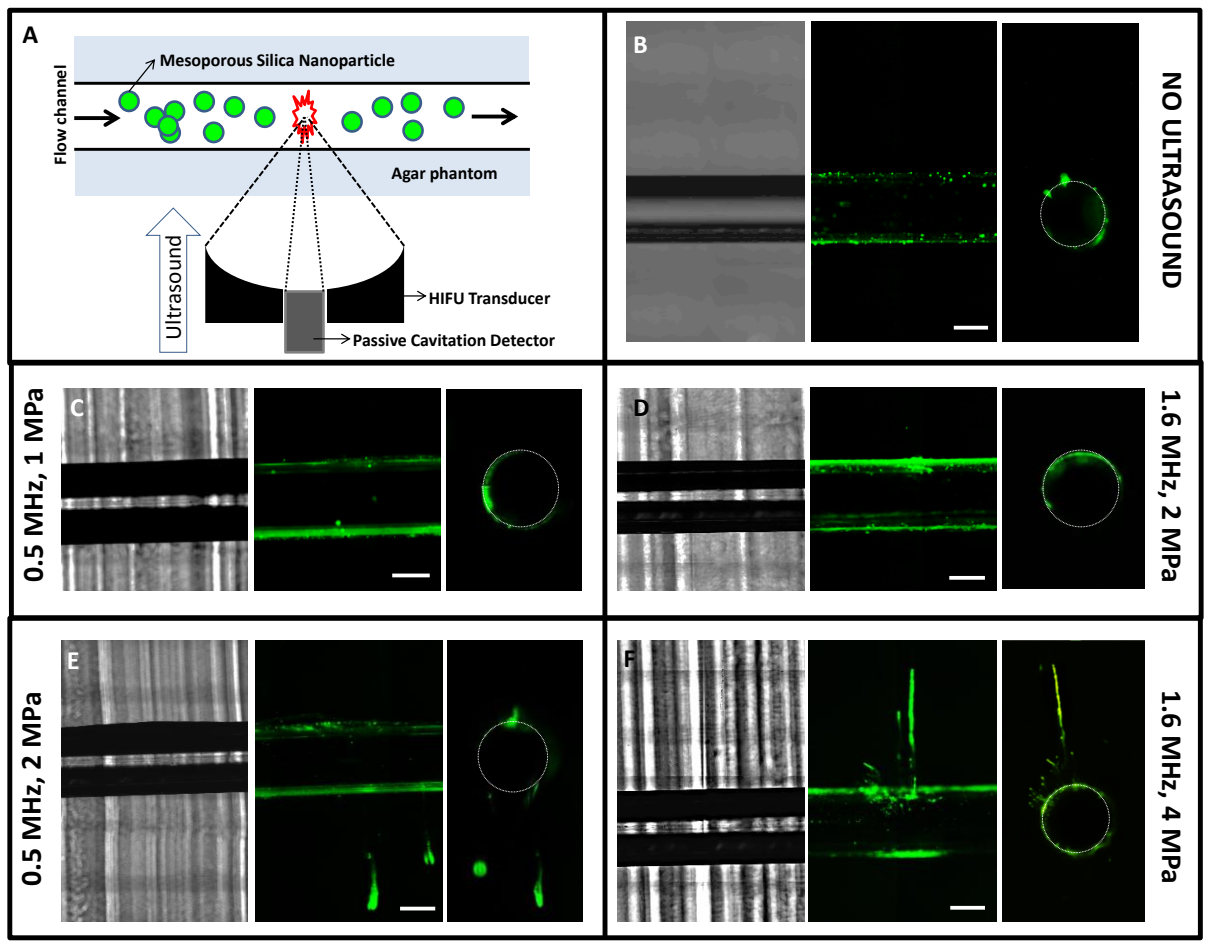

Figure 3.

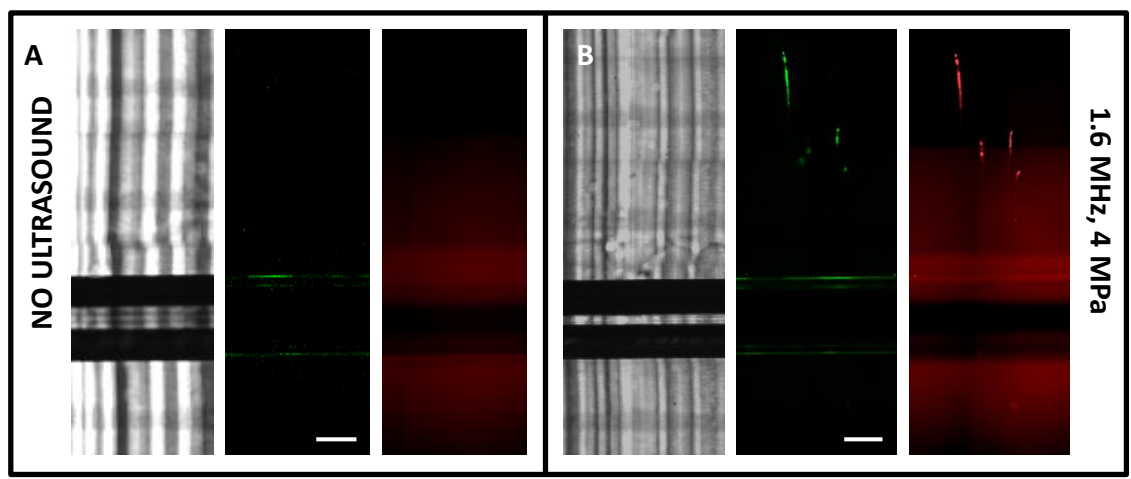

Figure 4.

**Figure 5**

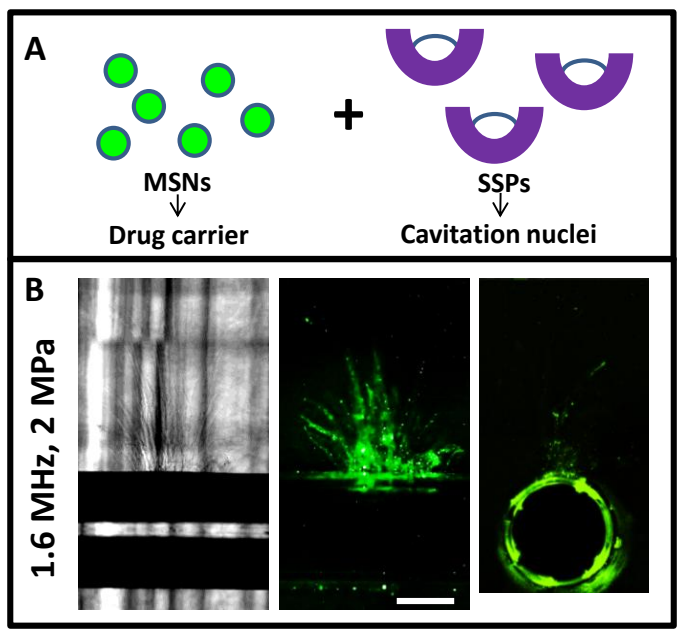

Figure 5.

**Figure 6**

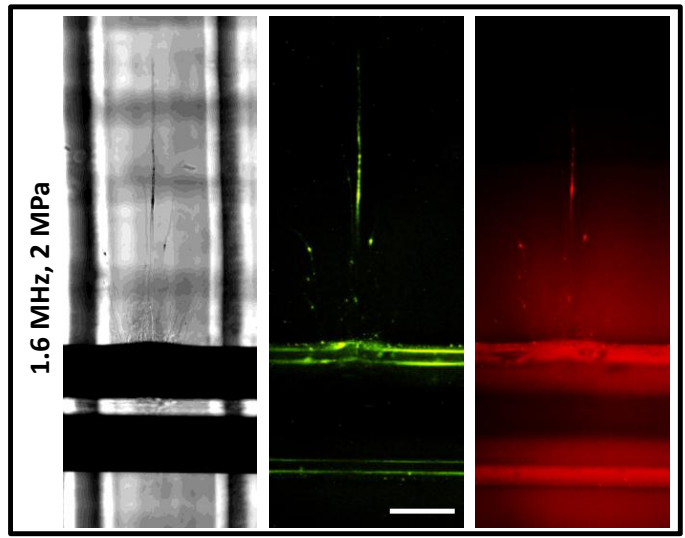

Figure 6.

**Figure captions:**

**Figure 1.** Schematic diagram of the experimental setup and the apparatus, containing the focussed ultrasound (FUS) transducer, passive cavitation detection (PCD) transducer, tissue-mimicking agarose phantom and diagnostic ultrasound imaging device (A), Photograph of the agarose phantom holder showing the gel with one of the three flow channels prepared (B) and B-mode images during experiment (C) showing displacement of air by incoming sample (top), channel with sample flowing (center) and inertial cavitation happening within the channel (bottom).

**Figure 2**. Probability of IC curves of FMSN in degassed water (200 µg/mL) exposed to ultrasound at 0.5 or 1.6 MHz. Data are Means ± SD, N=3.

**Figure 3.** Scheme of FMSN delivery experiments in the agarose phantom model (A). Representative microscopy images of Nanoparticle delivery in the agarose phantom model under different different frequency (0.5 or 1.6 MHz) and pressure (1-4 MPa) conditions at the same duty cycle (5%) (N=3), showing bright field (left), green fluorescence (center) and green fluorescence in a cross section of the flow channel (right) (B-F). Scale bars represent 500 µm.

**Figure 4.** Representative microscopy images of FMSN-RhB delivery in the agarose phantom model without ultrasound (A) or with FUS exposure at 1.6 MHz and 4 MPa (B), showing bright field (left), green fluorescence (center) and red fluorescence (right) (N=3). Scale bars represent 500 µm.

**Figure 5**. Schematic of combining MSNs with SSPs (A). Representative microscopy images of nanoparticle delivery in the agarose phantom model (FMSN in combination with SSPs), showing bright field (left), green fluorescence (center) and green fluorescence in a cross section of the flow channel (right) (B) (N=3). Scale bar represents 500 µm.

**Figure 6**. Representative microscopy images of dye-loaded nanoparticle delivery in the agarose phantom model (FMSN-RhB in combination with SSPs), showing bright field (left), green fluorescence (center) and red fluorescence (right) (B) (N=3). Scale bar represents 500 µm.